# Equivalence perspectives in communication, source-channel connections and universal source-channel separation

Mukul Agarwal, Anant Sahai, Sanjoy Mitter

**Abstract**

An operational perspective is used to understand the relationship between source and channel coding. This is based on a direct reduction of one problem to another that uses random coding (and hence common randomness) but unlike all prior work, *does not involve any functional computations, in particular, no mutual-information computations*. This result is then used to prove a universal source-channel separation theorem in the rate-distortion context where universality is in the sense of a compound "general channel."

## I. INTRODUCTION

The essential duality between source and channel coding has been recognized since Shannon [1] and has attracted significant attention recently as well (e.g. [2], [3], [4]). This paper addresses a conceptual issue: what is the core relationship between source and channel coding and to what extent do we need mutual-information computations to understand it?

Recall that classically, mutual information plays a critical role. After all, the traditional separation theorem (separate source and channel codes result in no loss in first-order[1] optimality when delay is not an issue.) relies crucially on the mutual-information characterization of both channel capacity and the rate-distortion function to prove the converse direction: that we can do no better. Even the more general framework of [2] builds upon the information-spectrum approach of [6] that extends mutual-information ideas to general channels by looking at the entire distribution of an information-random-variable instead of just the expectation.

Recently, a "direct" proof of the converse direction of the separation theorem was introduced by us in [7]. The key idea was to treat a combined joint-source-channel code as a non-causal arbitrarily-varying channel (AVC) with a particularly weak guarantee: as long as the input to it is drawn like the source in question, it will with high probability return an output within a specified distortion $D$. For such a channel, a random-coding argument revealed that reliable communication is possible at a rate given by the rate-distortion function of the source in question evaluated at $D$. The proof in [7], however, relied crucially on the mutual-information characterization of the rate-distortion function.

Conceptually, there are two distinct directions that one can explore from [7]. Lomnitz and Feder in [8] essentially emphasize the mutual-information aspects for a core result that avoids the need for an *a priori* distortion-guarantee and they then use feedback to translate this core into a meaningful interpretation concerning "communication over individual channels." This is in the spirit of individual-sequence results as distinct from the AVC perspective taken in [7]. The contribution of this work is to move in the complementary direction. After introducing notation and definitions in Section II, we give a new operational proof in Section III that *does not use mutual-information computations in any way*. It illuminates the operational connections and technical parallels between the problems of reliable communication at a particular rate and lossy-communication of a source to within a target distortion, in effect providing a direct "problem reduction" in the style that theoretical computer scientists use. It shows that the rate-distortion function of $X$ gives the universal capacity of the compound set of general channels that communicate i.i.d. $X$ sources to within a distortion $D$ (see Theorem 3.1 for a precise statement). *This naturally gives rise to a universal source-channel separation theorem in Section IV.*

## II. NOTATION AND DEFINITIONS

**Sets, random variables, and distortion measure:** Many symbols will have an interpretation for both rate-distortion source coding and channel coding problems. $\mathcal{X} = \{1, 2, \ldots, |\mathcal{X}|\} \to$ finite set should be thought of as the channel input alphabet or the alphabet of the source that needs to be source-coded. $\mathcal{Y} = \{1, 2, \ldots, |\mathcal{Y}|\}$ should similarly be thought of as the channel output alphabet or the reconstruction alphabet of the source. Let $X$ be a random variable on $\mathcal{X}$. $p_X$ will denote the corresponding 2probability distribution. $d : \mathcal{X} \times \mathcal{Y} \to \mathcal{R}$ is a non-negative real-valued function that represents the distortion incurred when $x \in \mathcal{X}$ is reconstructed as $y \in \mathcal{Y}$.

---

[1] Csiszar established in [5] that strict separation does result in a significant loss in the error-exponent. Separation also breaks down even in a first-order sense for multiterminal problems.

**Notation**: A superscript $n$ denotes a variable whose block length is $n$. For example, $Y^n$ will denote a random-variable on $\mathcal{Y}^n$.

**Method of Types:** We follow the notation of Csiszar and Korner [9].

**Channel model:** A channel is a sequence of transition-probability matrices and will be denoted by $<c^n>_1^\infty$. Its operation should be thought of as follows for block-length $n$: channel input space is $\mathcal{X}^n$, channel output space is $\mathcal{Y}^n$, and the channel acts as $c^n$: $c^n_{xy}$ is the probability that the channel output is $y \in \mathcal{Y}^n$ when channel input is $x \in \mathcal{X}^n$. No causality, memorylessness, or nestedness assumptions are assumed on $<c^n>_1^\infty$. This channel model is the same as that of Verdu and Han in [6].

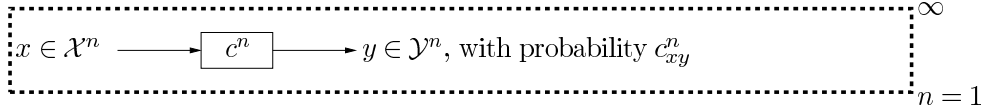

*Definition 2.1 ($\mathcal{C}_{X,D}$):* Consider a channel $<c^n>_1^\infty$. If the input to the channel is i.i.d. $X$ source $X^n$, the channel output is a *(not necessarily i.i.d.)* random variable $Y^n$ on $\mathcal{Y}^n$. A channel is said to belong to $\mathcal{C}_{X,D}$ if, under the joint distribution $p_{X^n Y^n}$ on the input-output space,

$$\Pr\left(\sum_{i=1}^n \frac{1}{n} d(X^n(i), Y^n(i)) > D\right) \to 0 \text{ as } n \to \infty \tag{1}$$

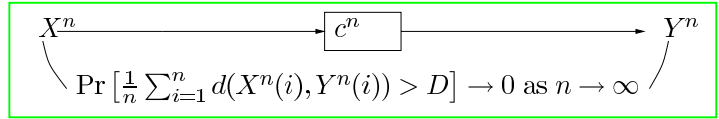

The i.i.d. $X$ sequence $X_i$ here is just a tool in the definition of the compound channel set $\mathcal{C}_{X,D}$. It does not mean that one is necessarily trying to communicate solely i.i.d. $X$ sources over the channel using uncoded transmission. Intuitively, one can think of a channel $\in \mathcal{C}_{X,D}$ as follows: most $p_X$-typical sequences of length $n$ are usually distorted to within a distortion $nD$. *Channels $\in \mathcal{C}_{X,D}$ will be called channels that directly communicate an i.i.d. $X$ source to within a distortion level $D$.*

**The process of communication:** Block codes will be used for communication with block length $n$. The channel input space $\mathcal{X}^n$, is the cartesian product of $\mathcal{X}$, $n$ times. $\mathcal{X}^n = \{x_1, x_2, \ldots, x_{|\mathcal{X}|^n}\}$. The channel output space $\mathcal{Y}^n$, is the cartesian product of $\mathcal{Y}^n$, $n$ times. $\mathcal{Y}^n = \{y_1, y_2, \ldots, y_{|\mathcal{Y}|^n}\}$. If we want to communicate at rate $R$, the message set is $\mathcal{M}^n = \{1, 2, \ldots, 2^{nR}\}$. The message reproduction set $\widehat{\mathcal{M}}^n$ is the same as $\mathcal{M}^n$. A *deterministic* encoder is a map $e^n : \mathcal{M}^n \to \mathcal{X}^n$ and similarly, a deterministic decoder is a map $d^n : \mathcal{Y}^n \to \widehat{\mathcal{M}}^n$. Deterministic encoder-decoders will be denoted as d-encoder-decoders. A *stochastic-coupled sc-encoder-decoder* is the same as a random code. The encoder comes from a family of codes and the decoder has access to the realization of the encoder through *common randomness* — that is the encoder and decoder have access to a shared random variable of sufficient entropy. We do not worry here about how much common randomness is used. For a given block length, stochastic-coupled encoder-decoders will be denoted by $(e^n, d^n)$ and overall by $(e, d) = <e^n, d^n>_1^\infty$.

**Universal capacity:** Consider a compound set of channels $\mathcal{A}$. Consider a uniform distribution $M^n$ on $\mathcal{M}^n$ so $p_{M^n}(m) = \frac{1}{2^{nR}} \forall m \in \mathcal{M}^n$. Each composition of the $M^n$, encoder, channel from $\mathcal{A}$ and decoder results in an output random variable $\widehat{M}^n$ on $\widehat{\mathcal{M}}^n$. This induces a joint probability distribution $p_{M^n \widehat{M}^n}$ on the message-message reproduction space $\mathcal{M}^n \times \widehat{\mathcal{M}}^n$. Rate $R$ is universally achievable over $\mathcal{A}$ under the average block error probability criterion if there exist encoder-decoder pairs such that under this joint probability distribution, $\Pr(\widehat{M}^n \neq M^n) \to 0$ as $n \to \infty$ for each channel in $\mathcal{A}$. The randomness of the message and the randomness in the encoder-decoder are presumed to be independent of the channel. *The supremum of achievable rates is called the universal channel capacity $C_{sc}(\mathcal{A})$.*

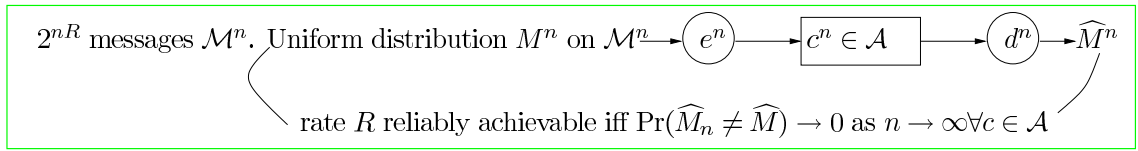

The channel set $\mathcal{A}$ can be interpreted as an adversary and in particular $\mathcal{C}_{X,D}$ is an adversary about which something specific is known. One can ask the question of universal capacity of $\mathcal{A}$ by restricting the set of encoders and decoders to be d or sc, and in general, one will get two different answers. *Error criteria different from $\Pr(\widehat{M}^n \neq M^n) \to 0$ as $n \to \infty$, also exist, but they will not be considered in this paper.*

**Source-code and operational rate-distortion function:** The source-coding problem is to code an i.i.d. $X$ source to within a distortion level $D$ in the sense of (1) while using the smallest rate possible to do so. The goal is to find a deterministic mapping whose output has the minimum cardinality and hence the smallest possible rate. See [9] for a precise statement. The minimum possible rate is called the operational rate-distortion function and is denoted by $R_X(D)$.

*The two problems between which we will see that there is a close connection:*

- *Universal Capacity of $\mathcal{C}_{X,D}$, and*
- *Source coding i.i.d. $X$ to within a distortion level $D$*

*That the set of all (potentially random) source-codes which code an i.i.d. $X$ source to within a distortion level $D$ is the same as $\mathcal{C}_{X,D}$ is the reason why these two questions are closely connected.*

## III. $C_{sc}(\mathcal{C}_{X,D}) = R_X(D)$: CONNECTION BETWEEN SOURCE AND CHANNEL-CODING

*Theorem 3.1:* $C_{sc}(\mathcal{C}_{X,D}) = R_X(D)$

Proof: First proved in [7]. We give another proof here that directly shows the close connections between the source-coding and channel-coding questions. The proof consists of two steps:

- A rate-distortion source-code can be interpreted as a particularly "bad" channel. The capacity of this "bad" channel is capped at $R^X(D)$ by a simple cardinality bound. Thus, $C_{sc}(\mathcal{C}_{X,D}) \leq R_X(D)$.
- There is a random coding-scheme for which rates $< \alpha$ are achievable for $\mathcal{C}_{X,D}$. Since there might be another scheme which performs even better, $C_{sc}(\mathcal{C}_{X,D}) \geq \alpha$.
  Similarly, there is a coding-scheme for which rates $> \alpha$ are achievable for the source-coding problem. There might be another scheme which performs even better and so $R_X(D) \leq \alpha$. Thus, $R_X(D) \leq \alpha \leq C_{sc}(\mathcal{C}_{X,D})$.

For $R_X(D) \geq C_{sc}(C_{X,D})$, only a little more detail is needed. Consider a "good" rate-$R_X(D)$ source-code. Now this source-code is a channel $\in \mathcal{C}_{X,D}$ with no more than $2^{nR_X(D)}$ possible outputs. Thus, the capacity of this channel $\leq R_X(D)$ because if we try to communicate at rate $> R_X(D)$, "many" codewords will get mapped to the same output sequence. This argument can be made precise (but longer) using standard techniques and proves $C_{sc}(\mathcal{C}_{X,D}) \leq R_X(D)$.

Next, we prove $R_X(D) \leq C_{sc}(\mathcal{C}_{X,D})$ using parallel random coding arguments, placing those for channel-coding and source-coding side by side to see the connection. See below:

| $C_{sc}(\mathcal{C}_{X,D})$ | $R_X(D)$ |
|---|---|
| Achievability | Achievability |
| Codebook generation: | Codebook generation: |
| Generate $2^{nR}$ codewords i.i.d. $p_X$ | Generate $2^{nR}$ codewords independently of each other, each with *precise type* $q_Y$ ($q_Y$ is some prob. distribution on $\mathcal{Y}$) |
| This is the codebook $\mathcal{K}$. Note: Codewords $\in \mathcal{X}^n$ | This is the codebook $\mathcal{L}$. Note: Codewords $\in \mathcal{Y}^n$ |
| Joint Typicality: Let $\epsilon > 0$. $(x,y)$ jointly typical if | Joint Typicality: Let $\epsilon > 0$. $(x,y)$ jointly typical if |
| i. $x$ $\epsilon$-typical: $p_x \in \mathcal{T}(p_X, \epsilon)$ | i. $x$ $\epsilon$-typical: $p_X \in T(p_X, \epsilon)$ |
| ii. $\frac{1}{n}\sum_{i=1}^{n} d(x(i), y(i)) \leq D$ | ii. $\frac{1}{n}\sum_{i=1}^{n} d(x(i), y(i)) \leq D$ |
|  | iii. $q_y = q_Y$ |
| $y$ is output of a channel $\in \mathcal{C}_{X,D}$ | $y$ is generated with precise type $q_Y$ |
| Thus, there is no restriction on $q_y$ or $q_{y|x}$ | Thus, iii above is redundant. |
| $x \in \mathcal{K}$ will denote transmitted codeword | $x \in \mathcal{X}^n$ will denote sequence to be source-coded |
| $y \in \mathcal{Y}^n$ will denote received sequence | $y \in \mathcal{L}$ will denote a codeword |
| $z \in \mathcal{K}$ will denote non-transmitted codeword |  |
| Decoding strategy | Encoding strategy |
| If $\exists$ *unique* $x \in \mathcal{K}$ such that $(x,y)$ jointly typical declare $x$ is transmitted. Else declare error. | If $\exists$ *some* $y \in \mathcal{L}$ such that $(x,y)$ jointly typical encoder $x$ to one such $y$. Else declare error. |
| Error events | Error events |
| $\mathcal{E}_1$: $(x,y)$ not $\epsilon$-jointly typical | $\mathcal{F}_1$: $x$ not $\epsilon$-typical |
| $\mathcal{E}_2$: $\exists z \neq x \in \mathcal{K}$ such that | $\mathcal{F}_2$: $\nexists y \in \mathcal{L}$ such that |
| $(z, y)$ $\epsilon$-jointly typical | $(x, y)$ $\epsilon$-jointly typical given $x$ $\epsilon$-jointly typical |
| $\Pr(\mathcal{E}_1) \to 0$ as $n \to \infty$ by $\mathcal{C}_{X,D}$ defn. | $\Pr(\mathcal{F}_1) \to 0$ as $n \to \infty$ by WLLN. |
| Analysis of $\Pr(\mathcal{E}_2)$: | Analysis of $\Pr(\mathcal{F}_2)$: |
| $z$ is generated i.i.d. $p_X$, | $y$ is generated with precise type $q_Y$ |
| independently of $y$ | independently of $x$ |
| The calculation required is the following: | The calculation required is the following: |
| fix type of $y$, the output type to be $q_Y$. | fix type of $y$, the output type to be $q_Y$. |
| Calculate probability that $(z, y)$ jointly typical | Calculate probability that $(z, y)$ jointly typical |
| given that $x$ is typical | given that $x$ is typical |
| Take *worst* case over $q_Y$ | Take *best* case over $q_Y$ |
| Worst case: maximize error probability | Best case: maximize probability that encoding is possible |
| Thus, as $\epsilon \to 0$, $q_Y$ for both problems is same | Thus, as $\epsilon \to 0$, $q_Y$ for both problems is same |
| Thus, answer to both calculations is same: call it $F(n)$ | Thus, answer to both calculations is same: call it $F(n)$ |
| Now, take a bound for whole codebook | Now, take a bound for whole codebook |
| If $(1 - F(n))^{2^{nR}} \to 1$ as $n \to \infty$, rate $R$ is achievable | If $(1 - F(n))^{2^{nR}} \to 0$ as $n \to \infty$, rate $R$ is achievable. |

Now, it turns out that $(1 - F(n))^{2^{nR}}$ exhibits a tight phase-transition as $n$ gets large. Make $R$ a little bigger and it goes to 0 and a little smaller and it goes to 1. It follows that there is a threshold $\alpha$ such that all rates $< \alpha$ are achievable for the channel-coding problem and all rates $> \alpha$ are achievable for the source-coding problem. Thus, $R_X(D) \le \alpha \le C_{sc}(\mathcal{C}_{X,D})$. ∎

*Notice that this argument does not have to do any calculations for either capacity or the rate-distortion function. We just use the operational definition of capacity as the maximum rate of reliable communication and the operational definition of the rate-distortion function as the minimum rate required to source-code $X$ to within a distortion $D$.*

## IV. Universal source-channel separation theorem for rate-distortion assuming common randomness

In this section, we prove a universal source-channel separation theorem in the rate-distortion context, *where universality is over the channel*. We also see an operational, direct view of source-channel separation for rate-distortion.

**Universal lossy communication to within a distortion $D$ over a channel set $\mathcal{A}$:** Channel set $\mathcal{A}$ is said to be capable of universally communicating an i.i.d. $X$ source to within a distortion level $D$ if there exist encoder-decoders $e = <e^n>_1^\infty, <d^n>_1^\infty$ such that all the composite channels (the composition of encoder, channel from $\mathcal{A}$, and decoder) $<d^n \circ c \circ e^n>_1^\infty$, directly communicate an i.i.d. $X$ source to within a distortion $D$, for all $c = <c^n>_1^\infty \in \mathcal{A}$. In other words, $d \circ c \circ e \in \mathcal{C}_{X,D}$ for all $c \in \mathcal{A}$.

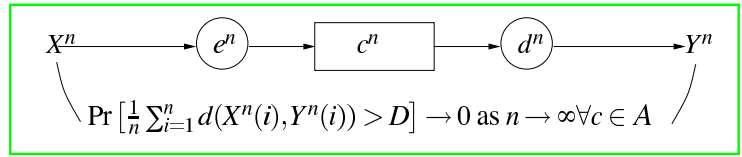

The composite channel set $\{d \circ c \circ e \, : \, c \in \mathcal{A}\}$ will be denoted by $d \circ \mathcal{A} \circ e$.

*Theorem 4.1 (Universal source-channel theorem for rate-distortion (USCS)):* Assuming there is common randomness, in order to communicate i.i.d. $X$ to within a distortion level $D$ universally over a channel set $\mathcal{A}$, it is sufficient to consider architectures which first source code the source to within a distortion level $D$ followed by universal reliable channel coding over $\mathcal{A}$.

Proof: Let $\mathcal{A}$ be a channel set. Consider the following three statements.

$S_1 : C_{sc}(\mathcal{A}) > R_X(D)$. $\quad S_1^* : C_{sc}(\mathcal{A}) \ge R_X(D)$. $\quad S_2 : \mathcal{A}$ is capable of universally communicating an i.i.d. $X$ source to within a distortion $D$ using an sc-encoder-decoder.

Proof of $S_1 \Rightarrow S_2$ is the usual argument of source-coding followed by channel-coding. Roughly, source-code the i.i.d. $X$ source. The output of the source-code is a message set of cardinality $2^{nR_X(D)}$ with a probability distribution on it. Communicate the message universally, reliably over $\mathcal{A}$ with an sc-encoder-decoder. This proof is rough, but since everything involved is standard, a precise proof is omitted. This completes the proof of $S_1 \Rightarrow S_2$.

To prove $S_2 \Rightarrow S_1^*$: We will keep referring to the figure below which gives a step by step view of the argument. $\mathcal{A}$ (black rectangle) is capable of universally communicating i.i.d. $X$ to within a distortion $D$ with an sc-encoder-decoder. That is, $\exists$ sc encoder-decoder $e_a = <e_a^n>_1^\infty, d_a = <d_a^n>_1^\infty$ such that $\forall c = <c^n>_1^\infty \in \mathcal{A}$, the composite channel $d_a \circ c \circ e_a = <d_a^n \circ c^n \circ e_a^n>_1^\infty$ directly communicates an i.i.d. $X$ source to within a distortion $D$ for all $c \in \mathcal{A}$. That is, $d_a \circ c \circ e_a \in \mathcal{C}_{X,D} \forall c \in \mathcal{A}$. More compact way of saying this is that channel set $\mathcal{C}_\mathcal{A} \triangleq d_a \circ \mathcal{A} \circ e_a \subset \mathcal{C}_{X,D}$ (yellow rectangle). Thus, by Theorem 3.1 $C_{sc}(\mathcal{C}_\mathcal{A}) \geq R_X(D)$. So there exists an sc-encoder-decoder $e_b = <e_b^n>_1^\infty, d_b = <d_b^n>_1^\infty$ such that with this encoder-decoder, there is reliable communication across $\mathcal{C}_\mathcal{A}$ (magenta rectangle). Now, $d_b \circ \mathcal{C}_\mathcal{A} \circ e_b = d_b \circ d_a \circ \mathcal{A} \circ e_a \circ e_b = (d_b \circ d_a) \circ \mathcal{A} \circ (e_b \circ e_a) = d_f \circ \mathcal{A} \circ e_f$ where $d_f \triangleq (d_b \circ d_a)$ and $e_f \triangleq (e_b \circ e_a)$ is an sc-encoder-decoder pair (red rectangles) that achieve universal reliable communication over $\mathcal{A}$. Thus, $C_{sc}(\mathcal{A}) \geq R_X(D)$. This proves $S_2 \Rightarrow S_1^*$. Theorem 4.1 follows ∎.

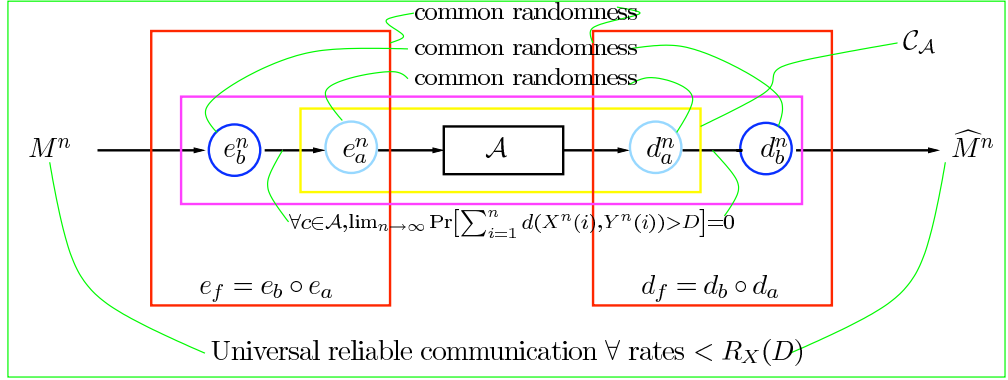

*One gets the following layered architecture for reliable communication: an architecture for reliable communication built "on top of" an architecture for communication to within a distortion $D$. See figure above. $d_a \circ c \circ e_a$ is the architecture for communication to within a distortion $D$ (blue rectangle), and an architecture for reliable communication, using encoder-decoder $e_b, d_b$ is built "on top of" it (magenta rectangle). This is a "direct" reduction of the reliable communication problem to the problem of 'communication to within a distortion $D$.' The view is operational because the proofs of Theorems 3.1 and 4.1 are operational. We believe that the USCS perspective might be useful in network problems.*

## V. Conclusion

The results in this paper imply that there are natural equivalence relationships among communication problems. Here, the equivalence is shown by explicit reductions from one problem to another in a spirit analogous to [10]. In light of this, the traditional mutual-information characterization of rate can be viewed as a kind of key "invariant" that labels the equivalence classes. The implication is that if common-randomness is available, then there is nothing sacred about the traditional layering: source-coding followed by reliable communication over channels. Instead, the inner layer could just as well be something that is only guaranteed to communicate e.g. an asymmetric ternary source with $P(a) = 2P(b) = 3P(c) = \frac{1}{6}$ to within Hamming distortion $\frac{1}{9}$. There will be no loss of optimality by forcing this seemingly bizarre architecture.

However, it turns out that the common-randomness is really critical: in general for any Theorem 3.1 style universal reduction of a reliable communication problem to one with non-zero distortion, a significant amount of common-randomness is required [11]. This suggests that there might be something special about the traditional layering after all: no additional common-randomness is required if the inner layer gives a reliable communication guarantee. Furthermore, it suggests that there might be other interesting "invariants" out there besides the rate-distortion function even for the simple stationary memoryless sources with additive distortion measures considered here.